\documentclass[aps,prb,twocolumn,groupedaddress,showpacs]{revtex4-1}
\usepackage{graphics}
\usepackage{graphicx}
\usepackage{amsmath}
\usepackage{amssymb}
\usepackage{amsthm}
\usepackage{epsfig}
\usepackage{dsfont}
\begin{document}
\title{On the physical significance of London's equation}
\author{Jacob Szeftel$^1$, Nicolas Sandeau$^2$, Antoine Khater$^3$}
\affiliation{$^1$ENS Cachan, LPQM, 61 avenue du Pr\'esident Wilson, 94230 Cachan, France}
\affiliation{$^2$Aix Marseille Universit\'e, CNRS, Centrale Marseille, Institut Fresnel UMR 7249, 13397, Marseille, France}
\affiliation{$^3$Institut des Mol\'ecules et Mat\'eriaux du Mans IMMM UMR 6283, LUNAM, Universit\'e du Maine, F-72000 Le Mans France}
\date{\today}
\begin{abstract}
The Meissner effect is analysed by using an approach based on Newton and Maxwell's equations, in order to assess the relevance of London's equation. The Hall effect is predicted. Two test experiments are proposed in detail to check the validity of this theory and to measure London's length.
\end{abstract}
\maketitle
		\section{introduction}
Superconductivity is characterized by two prominent properties\cite{ash,par}: persistent currents in vanishing electric field and the Meissner effect\cite{mei}, which highlights the rapid decay of an applied magnetic field within bulk matter in a superconductor of type I or II, provided the field is lower than the critical fields $H_c$ or $H_{c1}$, respectively. Some insight into the Meissner effect could be achieved  thanks to London's assumption\cite{lon}
\begin{equation}
\label{lon}
B+\mu_0\lambda^2_L\textrm{curl} j=0\quad,
\end{equation}
where $\mu_0,j,\lambda_L$ stand for the magnetic permeability of vacuum, the persistent current, induced by the magnetic induction $B$ and London's length, respectively. Eq.(\ref{lon}), combined with Newton and Maxwell's equations, entails\cite{ash,par,lon} that the penetration depth of the magnetic field is equal to $\lambda_L$, inferred\cite{ash,par,lon,gen} as
	$$\lambda_L=\sqrt{\frac{m}{\mu_0\rho e^2}}\quad,$$
where $e,\quad m,\quad\rho$ stand for the charge, effective mass and concentration of superconducting electrons.\par
	The first one to question the validity of Eq.(\ref{lon}) was Pippard\cite{pip1,pip} who investigated the effect of impurities on the absorption of electromagnetic waves at microwave frequencies in superconducting $Sn$ and favored a phenomenological interpretation, based on the anomalous skin effect\cite{reu,cha}. This has resulted in interesting but inconclusive debates, regarding the validity of Eq.(\ref{lon}) :
\begin{itemize}
	\item 
	some authors\cite{edw,gen,ess} have attempted to justify Eq.(\ref{lon}) by a classical treatment, whereas another school claimed that the Meissner effect stemmed from some unknown quantum effect\cite{yos}, possibly related to the BCS theory\cite{bar} and Cooper pairs\cite{coo}; 
		\item
		when a superconducting material is cooled in a magnetic field $H$, starting from its normal state, the latter is expelled\cite{mei,pip2} from the bulk material, while crossing the critical temperature $T_c(H)$ at which superconductivity sets in. This additional manifestation of the Meissner effect has generated an inconclusive debate over the distinction between a real material superconductor and a fictitious perfect conductor\cite{ash,par,lon,hen}.
\end{itemize}
	 However, nowadays all measurements of microwave energy absorption, carried out in superconducting materials\cite{h2,h1,gor,har,son}, are intended at assigning the skin depth\cite{bor}, which describes the finite penetration of the electromagnetic field. As the skin depth is $\propto 1/\sqrt{\sigma\omega}$, where $\sigma,\omega$ stand for the conductivity of superconducting electrons and the microwave frequency, it is widely accepted that all superconductors display a finite conductivity at $\omega\neq 0$, which however is consistent with the observation of persistent currents at vanishing electric field.\par
 	Meanwhile a recent work by Hirsch\cite{hir,hir2} deserves a special mention, because it seems to be the first one challenging the well-entrenched claim that the London-BCS theory\cite{lon,bar} accounts satisfactorily for the whole physics of the Meissner effect. It also makes a prediction, to be validated hereafter, that electron charge might pile up at the outer edge of a superconducting sample, embedded in a magnetic field. The present work takes advantage of Hirsch's study in order to work out a theory of the Meissner effect, resorting solely to classical tools, and to assess the validity of Eq.\ref{lon}.\par
  The outline is as follows: Sections II and III deal with the Meissner effect. The validity of London's assumption, expressed by in Eq.\ref{lon}, is analyzed in Section IV. The case of the field cooled superconductor is addressed in Section V. An experiment, enabling one to assess the validity of this theoretical approach is detailed in Section VI, and the Hall effect is dealt with in Section VII. The experimental measurement of $\lambda_L$ is described in Section VIII. The conclusions are given in Section IX.\par
\begin{figure}
\includegraphics*[height=5.5 cm,width=8 cm]{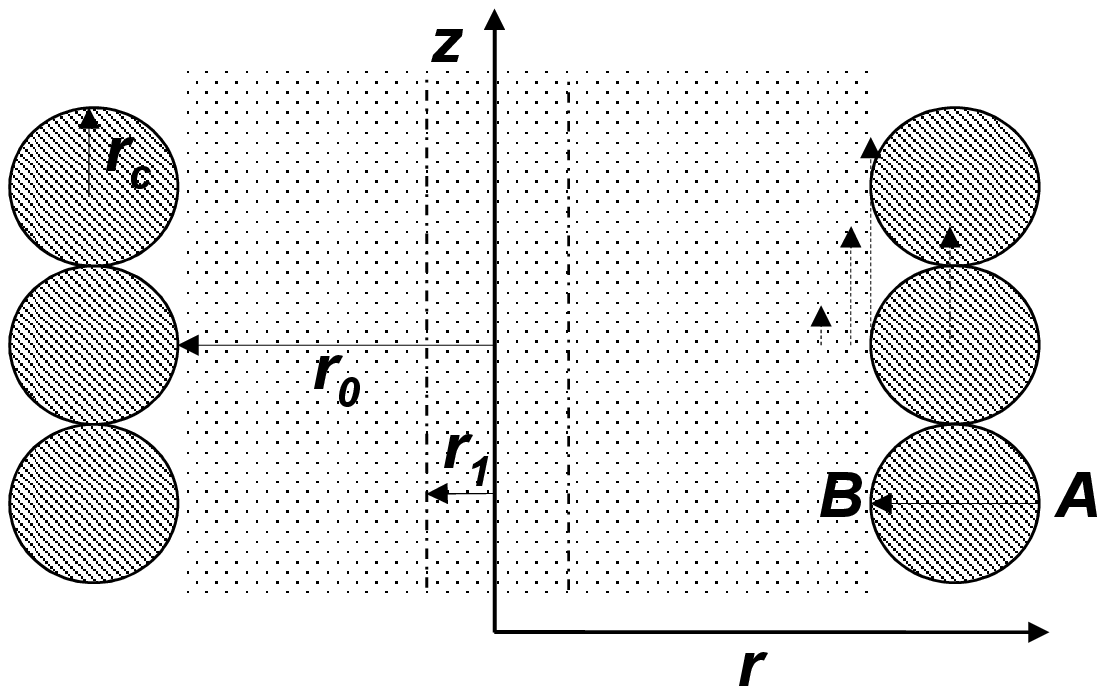}
\caption{Cross-section of the superconducting sample (dotted) and the coil (hatched); $E_\theta$ and $j_\theta$ are both normal to the unit vectors along the $r$ and $z$ coordinates; vertical arrows illustrate the $r$ dependence of $B_z(r)$; $r_c$ has been magnified for the reader's convenience; Eq.(\ref{coil}) has been integrated from $A$ ($B_z(r_0+2r_c)=0$) to $B$; the matter between the dashed-dotted lines should be carved out to carry out the Hall effect experiment}\label{Bzr}
\end{figure}
	Consider as in Fig.1 a superconducting material of cylindrical shape, characterized by its symmetry axis $z$ and radius $r_0$ in a cylindrical frame with coordinates ($r,\theta,z$). The material contains superconducting electrons of charge $e$, effective mass $m$ and concentration $\rho$. It is subjected to a time $t$ dependent electric field $E_\theta(t,r)\neq 0$ only during $t\in ]0,t_0[$, which defines a transient regime $(0<t<t_0)$ and a permanent one $(t>t_0)$. As $E_\theta(t,r)$ is normal to the unit vectors along the $r$ and $z$ coordinates, there is $\textrm{div} E_\theta=0$.
		\section{transient regime}
$E_\theta$ induces a current $j_\theta(t,r)$ along the field direction, as given by Newton's law
\begin{equation}
\label{newt}
\frac{dj_\theta}{dt}=\frac{\rho e^2}{m}E_\theta-\frac{j_\theta}{\tau}\quad,
\end{equation}
where $\frac{\rho e^2}{m}E_\theta$ and $-\frac{j_\theta}{\tau}$ are respectively proportional to the driving force accelerating the conduction electrons and a generalized friction term, which does not vanish in a superconductor, only if $\frac{dj_\theta}{dt}\neq0$. However the physical sense of $\tau$ in Eq.(\ref{newt}) for superconductors may be quite different from that given by the Drude model for a normal metal\cite{ash}. To understand this difference and to account for the new $\tau$, we shall next work out the equivalent of Ohm's law for a superconductor, submitted to an electric field.\par
The superconducting state, carrying no current, is assumed to comprise two subsets of equal concentration $\rho/2$, moving in opposite directions with respective mass center velocity $v,-v$, which ensures $j_\theta =p=0$, where $p$ refers to the average electron momentum. The driving field $E_\theta$ causes $\delta\rho/2$ of electrons to be transferred from one subset to the other, so as to give rise to a finite current $j_\theta=\delta\rho ev=e\delta p/m$, where $\delta p$ stands for the electron momentum variation. The generalized friction force is responsible for the reverse mechanism, whereby electrons are transferred from the majority subset of concentration $\frac{\rho+\delta\rho}{2}$ back to the minority one ($\frac{\rho-\delta\rho}{2}$). It ensues from flux quantization and Josephson's effect\cite{ash,par,jos} that the elementary transfer process involves a pair rather than a single electron. Hence if $\tau^{-1}$ is defined as the transfer probability per unit time of one electron pair, the net electron transfer rate is equal to $\frac{\rho+\delta\rho-(\rho-\delta\rho)}{2\tau}=\frac{\delta\rho}{\tau}$. By virtue of Newton's law, the resulting generalized friction term is equal to $mv\delta\rho/\tau=\delta p/\tau\propto j_\theta/\tau$, which validates Eq.(\ref{newt}), and permits to retrieve from it the equivalent of Ohm's law for the superconducting state as
		$$j_\theta=\sigma E_\theta\quad,\quad\sigma=\frac{\rho e^2\tau}{m}\quad,$$
whenever the inertial term $\propto\frac{dj_\theta}{dt}$ in Eq.(\ref{newt}) is negligible.\par
Although the conductivity $\sigma$ for the superconducting state has the same form as for the normal state\cite{ash}, its value has been found\cite{h2,h1,gor,har} to be $\approx300$ times greater.\par
$E_\theta$ induces a magnetic induction $B_z(r,t)$, parallel to the $z$ axis. $B_z$ is given by the first Maxwell equation as
\begin{equation}
\label{Bz}
\frac{\partial B_z}{\partial t}=-\textrm{curl}E_\theta=-\left(\frac{E_\theta}{r}+\frac{\partial E_\theta}{\partial r}\right)\quad.
\end{equation}
The displacement vector $D$, is parallel to $E_\theta$ and is defined as
	$$D=\epsilon_0 E_\theta+\rho e u_\theta\quad,$$
where $\epsilon_0,\quad u_\theta$ refer to the electric permittivity of vacuum and displacement coordinate of the conduction electron center of mass, parallel to $E_\theta$. The term $\rho e u_\theta$ represents the electric polarization of conduction electrons \cite{not2}. Because $\textrm{div} E_\theta=0$ entails that $\textrm{div} D_\theta=0$, Poisson's law warrants the lack of charge fluctuation around $\rho e$. Thence since there is by definition $j_\theta=\rho e\frac{du_\theta}{dt}$, the displacement current reads 
	$$\frac{\partial D_\theta}{\partial t}=j_\theta+\epsilon_0\frac{\partial E_\theta}{\partial t}\quad.$$
Finally the magnetic field $H_z(t,r)$, parallel to the $z$ axis, is given by the second Maxwell equation as
\begin{equation}
\label{Hz}
\textrm{curl}H_z=-\frac{\partial H_z}{\partial r}=j_\theta+\frac{\partial D_\theta}{\partial t}=2j_\theta+\epsilon_0\frac{\partial E_\theta}{\partial t} \quad.
\end{equation}
$E_\theta\left(t,r\right),j_\theta\left(t,r\right),B_z\left(t,r\right),H_z\left(t,r\right)$ can be recast as Fourier series for $t\in ]0,t_0[$
\begin{equation}
\label{fou}
f\left(t,r\right)=\sum_{n\in\mathds{Z}} f\left(n,r\right)e^{in\omega_0 t}\quad,
\end{equation}
where $\omega_0 t_0=2\pi$ and $f\left(t,r\right),f\left(n,r\right)$ hold for  $B_z\left(t,r\right)$, $H_z\left(t,r\right)$, $E_\theta\left(t,r\right)$, $j_\theta\left(t,r\right)$ and $B_z\left(n,r\right)$, $H_z\left(n,r\right)$, $E_\theta\left(n,r\right)$, $j_\theta\left(n,r\right)$, respectively. Replacing $E_\theta,j_\theta,B_z,H_z$ in Eqs.(\ref{newt},\ref{Bz},\ref{Hz}) by their expression in Eqs.(\ref{fou}), while taking into account
	$$B_z\left(n,r\right)=\mu\left(n\omega_0\right)H_z\left(n,r\right)\quad,$$
where $\mu\left(n\omega_0\right)=\mu_0\left(1+\chi_s\left(n\omega_0\right)\right)$ and $\chi_s\left(\omega\right)$ is the magnetic susceptibility of superconducting electrons at frequency $\omega$, yields for $n\neq 0$
\begin{equation}\label{fou2}
\begin{array}{l}
E_\theta\left(n,r\right)=\frac{1+in\omega_0\tau}{\sigma}j_\theta\left(n,r\right)\\
in\omega_0 B_z\left(n,r\right)=-\left(\frac{E_\theta\left(n,r\right)}{r}+\frac{\partial E_\theta\left(n,r\right)}{\partial r}\right)\\
\frac{\partial B_z\left(n,r\right)}{\partial r}=-\mu\left(n\omega_0\right)\left(2j_\theta\left(n,r\right)+in\omega_0\epsilon_0E_\theta\left(n,r\right)\right)
\end{array}
\end{equation}
Eliminating $E_\theta\left(n,r\right)$ from Eqs.(\ref{fou2}) gives
\begin{equation}
	\label{skin}
\frac{\partial^2 B_z\left(n,r\right)}{\partial r^2}=\frac{B_z\left(n,r\right)}{\delta^2(n\omega_0)}-\frac{\partial B_z\left(n,r\right)}{r\partial r}\quad.
\end{equation}
$$\delta(\omega)=\frac{\lambda_L}{\sqrt{\left(1+\chi_s\left(\omega\right)\right)\left(\frac{2i\omega\tau}{1+i\omega\tau}-\frac{\omega^2}{\omega^2_p}\right)}}\quad,\quad\omega_p=\sqrt{\frac{\rho e^2}{\epsilon_0m}}$$
refer to skin depth and plasma frequency\cite{ash,bor}, respectively. As Eqs.(\ref{fou2}) make up a system of $3$ linear equations in terms of $3$ unknowns $j_\theta,E_\theta,B_z$, there is a single solution, embodied by Eq.(\ref{skin}).\par
\begin{figure}
\includegraphics*[height=8 cm,width=8 cm]{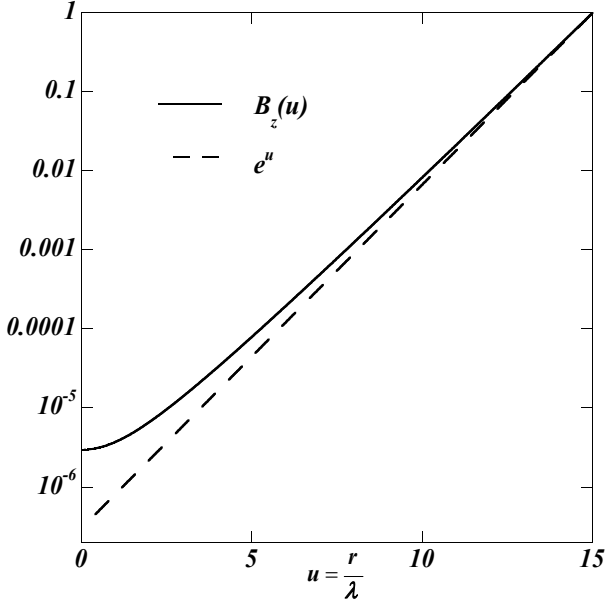}
\caption{Semi-logarithmic plots of $B_z(u),e^u$.}\label{fff}
\end{figure}
The solution of Eq.(\ref{skin}), which has been integrated over $r\in \left[0,r_0\right]$ with the initial condition $\dfrac{dB_z}{dr}\left(r=0\right)=0$, is a Bessel function, having the property $B_z(r)\approx e^{r/\delta(n\omega_0)}$ if $r>>|\delta(n\omega_0)|$, as illustrated in Fig.\ref{fff}. Finally, Eqs.(\ref{fou2}) entail that $E_\theta\left(n=0,r\right)=0\Rightarrow j_\theta\left(n=0,r\right)=0$, which in turn results into $H_z(n=0,r)=H_z(n=0,r_0),\forall r\in \left[0,r_0\right]$.\par
		\section{permanent regime}
Because $E_\theta(t>t_0,r)=0$ in the permanent regime, the generalized friction force $\propto-\frac{j_\theta}{\tau}$ is no longer at work for $t>t_0$, so that the \textit{transient} current $j_\theta(t< t_0)$ turns into the \textit{persistent} one, $j_\theta(t> t_0,r)=j_\theta(t_0,r),\forall r$. Eqs.(\ref{fou}) then yield
\begin{equation}
\label{jr}
j_\theta(t>t_0,r)=\sum_{n\in\mathds{Z}} j_\theta\left(n,r\right)\quad.
\end{equation}
  The second Maxwell equation reads now
\begin{equation}
\label{max2}
-\frac{\partial H_z}{\partial r}(t> t_0,r)=j_\theta(t_0,r) \quad.
\end{equation}
Comparing Eqs.(\ref{Hz},\ref{max2}) reveals that $H_z(t_{0^-},r)\neq H_z(t_{0^+},r)$. The penetration depth $\lambda_M$ is defined as
 $$\frac{1}{\lambda_M}=\frac{\partial \textrm{Log}H_z(t_{0^+},r_0)}{\partial r}\quad .$$
 At low frequencies such that $\omega\tau<<1$, one has $|\delta|\approx\lambda_L/\sqrt{\omega\tau}$. Given that $\lambda_L<10^{-7}$m, the inequality $r_0>>|\delta(n\omega_0)|$ holds for any $n$ under typical experimental conditions $r_0\approx 1$mm, $\omega_0<10^5Hz$. Using Eq.(\ref{jr}) and $j_\theta(n,r\rightarrow r_0)\approx j_\theta(n,r_0)e^{\frac{r-r_0}{\delta(n\omega_0)}}$ to integrate Eq.(\ref{max2}), we obtain
\begin{equation}
\label{mei2}
\frac{1}{\lambda_M}\approx\frac{\sum_n  j_\theta(n,r_0)}{\sum_n\delta(n\omega_0)j_\theta(n,r_0)-H_z(n=0,r_0)} \quad,
\end{equation}
where the sum is performed for $n\neq 0$ and $|n|\omega_0<\omega_p$. Thanks to Eq.(\ref{mei2}) and the inequality $|\delta(n\omega_0)|>>\lambda_L$, valid for $n$ such that $|n|\omega_0\tau<<1$, $|\lambda_M|$ is likely to be much larger than $\lambda_L$, which is well documented in experimental data\cite{pip1,pip,h2,h1,gor,har,son}. It is important to note that unlike $\lambda_L$, the $\lambda_M$ length depends on experimental conditions via $\omega_0$ and $j_\theta(n,r_0)$'s, and is hence not an intrinsic property of a superconductor.
	\section{validity of London's equation}
	Eq.(\ref{lon}) was assumed\cite{lon}, starting from the following version of Newton's equation
\begin{equation}
\label{newt1}
\frac{dj_\theta}{dt}=\frac{\rho e^2}{m}E_\theta\quad,
\end{equation}
which is identical to Eq.(\ref{newt}) in case $\tau \rightarrow\infty$. Integrating both sides of Eq.(\ref{newt1}) from $t=0$ up to $t=t_0$ yields for for $r\in [0,r_0]$
\begin{equation}
\label{lon1}
j_\theta(t_0,r)=\frac{\rho e^2}{m}\int_0^{t_0}E_\theta(t,r)dt=-\frac{\rho e^2}{m}A_\theta(t_0,r)\quad,
\end{equation}
by assuming $j_\theta(t=0,r)=A_\theta(t=0,r)=0$ and taking advantage of $E_\theta=-\frac{\partial A_\theta}{\partial t}$, where  the magnetic vector potential\cite{jac} $A_\theta(t,r)$ is parallel to $E_\theta$. Using furthermore $B_z=\textrm{curl} A_\theta$, it is inferred from Eq.(\ref{lon1}) for $r\in [0,r_0]$ in the permanent regime $t>t_0$
\begin{equation}
\label{lon2}
B_z+\mu_0\lambda^2_L\textrm{curl} j_\theta=0\quad,
\end{equation}
which is identical to Eq.(\ref{lon}). It has thereby been shown that London's equation is valid in the limiting case $\tau\rightarrow\infty$, which entails moreover that the penetration depth $\lambda_M=\lambda_L/\sqrt{2}$ is $\omega$ independent. However the measured\cite{h2,h1,gor,har,son} skin depth $\delta(\omega)$, being indeed $\propto 1/\sqrt{\omega}$, as expected theoretically\cite{bor}, confirms that $\tau$ is finite for $\omega\neq0$. 
	\section{field cooled sample}
	The expression of $\chi_s$ is needed for Eqs.(\ref{fou2}) to be self-contained and because the susceptibility not being continuous at $T_c$ will turn out to be \textit{solely} responsible for the Meissner effect to occur in a superconductor, cooled inside a magnetic field. Since no paramagnetic contribution has ever been observed in the superconducting state\cite{ash,par}, it has been concluded that the latter is always in a macroscopic singlet spin state. Consequently the only contribution to $\chi_s$ is of macroscopic origin and can thence be calculated using Maxwell's equations. We begin with writing down the $t$-averaged density of kinetic energy
	$$\mathcal{E}_K(r)=\frac{m}{2\rho}\left(\frac{j_\theta(r)}{e}\right)^2\quad,$$
associated with the current $j_\theta(r)e^{i\omega t}$, flowing along the $E_\theta$ direction (this latter induces in turn a magnetic field $H_z(r)e^{i\omega t}$, parallel to the $z$ axis). The second Maxwell equation simplifies into $\frac{\partial H_z}{\partial r}=-2j_\theta$ because the term $\propto E_\theta$ in the third equation in Eqs.(\ref{fou2}) shows up  negligible with respect to that $\propto j_\theta$ for practical $\omega<<\omega_p$. As this discussion is limited to the case $r\rightarrow r_0$, both $H_z(r),j_\theta(r)$ are $\propto e^{r/\delta(\omega)}$, so that $\mathcal{E}_K(r)$ is recast into
\begin{equation}
\label{k2}
\mathcal{E}_K(r)=\frac{\mu_0}{8}\left(\frac{\lambda_L}{|\delta(\omega)|}H_z(r)\right)^2\quad.
\end{equation}
Moreover there is the identity $\frac{\partial \mathcal{E}_K}{\partial M}=-H_z$, where $M=\mu_0\chi_s(\omega)H_z$ is the magnetization of superconducting electrons. Actually this identity reads in general $\frac{\partial F}{\partial M}=-H_z$, where $F$ represents the Helmholz free energy\cite{lan}; however the property that a superconducting state carries no entropy\cite{ash,par} entails that $F=E_K$. Equating this expression of $\frac{\partial \mathcal{E}_K}{\partial M}$ with that inferred from Eq.(\ref{k2}) yields finally
	$$\chi_s(\omega)=-\left(\frac{\lambda_L}{2|\delta(\omega)|}\right)^2\quad.$$
As expected, $\chi_s$ is found diamagnetic $(\chi_s<0)$ and $|\chi_s(\omega)|<<1$ for $\omega<<1/\tau$. The calculation of $\chi_s(0)$ proceeds along the same lines, except for the second Maxwell equation reading $\frac{\partial H_z}{\partial r}=-j_\theta$ and $\lambda_M$ showing up instead of $\delta(\omega)$, whence
	$$\chi_s(0)=-\left(\frac{\lambda_L}{\lambda_M}\right)^2\quad.$$
 Note that our definition of $\chi_s=\frac{M(r)}{\mu_0H_z(r)}$, where $H_z(r),M(r)$ refer to local field and magnetization at $r$, differs from the usual\cite{ash,par,lon,gen} one $\chi_s=\frac{M}{\mu_0H_z(r_0)}$ with $H_z(r_0),M$ being external field and total magnetization.\par
 While the sample is in its normal state at $T>T_c$, the applied magnetic field $H_z$ penetrates fully into bulk matter and induces a magnetic induction
\begin{equation}
\label{Bn}
B_n=\mu_0\left(1+\chi_n\right)H_z\quad,
\end{equation}
where $\chi_n$ designates the magnetic susceptibility of conduction electrons. It comprises\cite{ash} the sum of a paramagnetic (Pauli) component and a diamagnetic (Landau) one and $\chi_n>0$ in general. Moreover the magnetic induction reads for $T<T_c(H_z)$
\begin{equation}
\label{Bs}
B_s=\mu_0\left(1+\chi_s(0)\right)H_z\quad,
\end{equation}
with $\chi_s(0)<0$. Because of $\chi_s(0)\neq\chi_n$, the magnetic induction undergoes a finite step while crossing $T_c(H_z)$
\begin{equation}
\label{dBt}
\frac{\delta B}{\delta t}=\frac{B_s-B_n}{\delta t}=\mu_0\frac{\chi_s(0)-\chi_n}{\delta t}H_z\quad,
\end{equation}
where $\delta t$ refers to the time needed in the experimental procedure for $T$ to cross $T_c(H_z)$. Due to the first Maxwell equation (see Eq.(\ref{Bz})), the finite $\delta B/\delta t$ induces an electric field $E_\theta$ such that
$\textrm{curl} E_\theta=-\frac{\delta B}{\delta t}$, giving rise eventually to the persistent, $H_z$ screening current,  typical of the Meissner effect, as detailed hereabove.\par
	Noteworthy is that, though $H_z$ remains \textit{unaltered} during the cooling process, the magnetic induction $B$ is indeed \textit{modified} at $T_c$, as shown by Eq.(\ref{dBt}). This $B$ variation arouses the driving force, giving rise to the screening current $j_\theta$, and ultimately to $H_z$ expulsion, in accordance with Newton and Maxwell's law, as shown by Eq.(\ref{newt}) and Eq.(\ref{max2}), respectively.
	\section{test experiment}
An experiment, enabling one to check the validity of this work, will be presented now. It consists of inserting the superconducting sample into a cylindrical coil of radius $r_0$, flown through by an oscillating current $I_0(\omega)e^{i\omega t}$. The coil is made up of a wire of length $l$ and radius $r_c$ (see Fig.\ref{Bzr}). Applying Ohm's law to the coil yields
\begin{equation}
\label{ohm}
\begin{array}{l}
-l\left(E_a(\omega)+E_\theta(\omega,r_0)\right)=RI_0(\omega)\Rightarrow\\
E_\theta(\omega,r_0)=\frac{U_s(\omega)-RI_0(\omega)}{l}
\end{array}\quad,
\end{equation}
where $E_a(\omega)e^{i\omega t},E_\theta(\omega,r)e^{i\omega t},U_se^{i\omega t}=-lE_a(\omega)e^{i\omega t},R$ are the applied and induced electric fields, both normal to the $r,z$ axes, the voltage drop throughout the coil and its resistance, respectively $\left(E_a(\omega),E_\theta(\omega,r),U_s(\omega)\in\mathds{C}\right)$. Besides $E_\theta(\omega,r_0)$ is obtained from Eq.(\ref{fou2}) as
\begin{equation}
\label{eth}
E_\theta(\omega,r_0)=-i\omega\delta(\omega)B_z(\omega,r_0)\quad.
\end{equation}
where $E_\theta(\omega,r\rightarrow r_0)\approx E_\theta(\omega,r_0)e^{\frac{r-r_0}{\delta(\omega)}}$. Working out $B_z(\omega,r_0)$ in Eq.(\ref{eth}) requires to solve the second Maxwell equation for $B_z(\omega,r)e^{i\omega t}$ inside a cross-section of the coil wire
\begin{equation}
	\label{coil}
\frac{\partial B_z}{\partial r}=-\mu_0\left(2j_c+i\epsilon_0\omega E_c\right) \quad,	
\end{equation}
where $j_c(\omega)=\frac{I_0(\omega)}{\pi r_c^2}$ and $E_c(\omega)=E_a(\omega)+E_\theta(\omega,r_0)$ are both assumed to be $r$-independent. Moreover integrating Eq.(\ref{coil}) for $r\in\left[r_0+2r_c,r_0\right]$ with the boundary condition $B_z(\omega,r_0+2r_c)=0,\forall t$ (see Fig.\ref{Bzr}), while taking advantage of Eq.(\ref{ohm}), yields
\begin{equation}
\label{Bzt}
B_z(\omega,r_0)=2\mu_0\left(\frac{2}{\pi r_c}-\frac{i\epsilon_0\omega r_cR}{l}\right)I_0(\omega)\quad.
\end{equation}
Combining Eqs.(\ref{ohm},\ref{eth},\ref{Bzt}) leads finally to
$$\frac{R-U_s(\omega)/I_0(\omega)}{2\mu_0\omega l\delta(\omega)}=\frac{\epsilon_0\omega r_cR}{l}+\frac{2i}{\pi r_c}\quad.$$
Inserting the measured value of $U_s(\omega)/I_0(\omega)$ into that equation and checking that it is fulfilled for any $\omega$, would eventually ensure the validity of this analysis. In addition Eq.(\ref{eth}) predicts that
$\frac{U_n(\omega)-RI_0(\omega)}{U_s(\omega)-RI_0(\omega)}=\sqrt{r}$, where $U_n(\omega),r$ are the voltage drop amplitude, measured in the normal state, and the ratio of conductivities\cite{h2,h1,gor,har} pertaining to the superconducting and normal state, respectively.
	\section{the Hall effect}
	As already noted by Hirsch\cite{hir,hir2}, during the transient regime $t<t_0$, the magnetic induction $B_z$ exerts on the conduction electrons a radial Lorentz force $\frac{B_zj_\theta}{\rho}$, pushing the electrons outward, so that a charge distribution builds up, which in turn gives rise, via Poisson's law, to a radial electric field $E_r(r)$, typical of the Hall effect. It is noticeable that this Lorentz force arouses also a transient radial current but the latter, responsible for the charge distribution building up, vanishes in the permanent regime $t>t_0$, and is thence irrelevant to the Meissner effect.\par
	 Moreover for $t>t_0$, equilibrium is secured by the radial centrifugal force $\frac{m}{r}\left(\frac{j_\theta(t_0,r)}{\rho e}\right)^2$, exerted on each electron making up the persistent current $j_\theta(t_0,r)$, being counterbalanced by the sum of the Lorentz force and an electrostatic one $eE_r(r)$, with $E_r$ given by
	$$E_r=-\frac{j_\theta}{\rho e}\left(B_z+\frac{m}{\rho e^2r}j_\theta\right)\quad.$$
Owing to the second Maxwell equation $j_\theta=-\frac{\partial B_z}{\mu_0\partial r}$, $E_r$ can be recast as
	$$E_r=\frac{\partial B_z}{\partial r}\frac{B_z-\frac{\lambda_L^2}{r}\frac{\partial B_z}{\partial r}}{\mu_0\rho e}\quad.$$
 Because of $\frac{\partial B_z}{\partial r}\approx \frac{B_z}{\lambda_M}$, $r_0>>\lambda_L$ and $\lambda_M>>\lambda_L$, the approximation $E_r(r)\approx \frac{\partial B_z^2}{\partial r}/\left(2\mu_0\rho e\right)$ can be used for significant $r>>\lambda_L$. For the Hall effect to be observed, a sample in shape of a cylindrical crown of inner and outer radius $r_1,r_0$, respectively, is needed (see Fig.\ref{Bzr}). Finally the Hall voltage reads, for $r_0-r_1>>\lambda_M$
$$U_H=-\int_{r_1}^{r_0}E_r(r)dr\approx -\frac{B_z^2\left(t_0^+,r_0\right)}{2\mu_0\rho e}\quad,$$
 $B_z\left(t_0^+,r_0\right)=-\frac{2\mu_0}{\pi r_c}I(t_0)$ ($I(t_0)$ is the static current flowing through the coil for $t\geq t_0$) is worked out by integrating Eq.(\ref{max2}) under the same conditions used to integrate Eq.(\ref{coil}). As in normal metals\cite{ash}, measuring $U_H$ gives access to $\rho$. Note that $U_H$ is independent of $r_1$.\par
 	Moreover Poisson's law implies that a bulk charge density $\delta\rho$ piles up, which reads
 	$$\delta\rho(r)=\epsilon_0\textrm{div}E_r=\epsilon_0\left(\frac{E_r}{r}+\frac{\partial E_r}{\partial r}\right)\quad.$$
 This result validates Hirsch's prediction\cite{hir,hir2}. Finally charge conservation requires a further superficial charge density $\rho_S$ to build up at $r_0$ all over the outer surface of the sample  
 	$$\rho_S=-2\pi\int_0^{r_0}\delta\rho(r)dr\quad.$$

	\section{measurement of $\lambda_L$}
Most experiments\cite{pip1,pip,h2,h1,gor,har} have consisted of measuring complex impedances at frequencies $\omega\in\left[10MHz,30GHz\right]$, which is tantamount to assessing $\delta(\omega)$. Because of $\omega\tau<<1$ in that frequency range, there is $|\delta|\approx\lambda_L/\sqrt{2\omega\tau}=1/\sqrt{2\mu_0\sigma\omega}$. However whereas $\sigma$ can be measured by several methods, there is no experimental way to determine $\tau$, so that the exact value of $\lambda_L$ is not known and thence nor that of $\rho/m$.\par
 Therefore it is suggested to work at higher frequencies, such that $\omega>>1/\tau,\omega<<\omega_p$, because $\delta(\omega)=\lambda_L/\sqrt{2}$ is independent from $\tau$ in that range. Typical values $\tau\approx 10^{-11}s,\omega_p\approx 10^{16}Hz$ would imply to measure light absorption in the IR range. Then for an incoming beam being shone at normal incidence on a superconductor of refractive index $\tilde{n}\in\mathds{C}$, the absorption and reflection coefficients $A,R$ read\cite{bor}
  	$$A=1-R=1-\left|\frac{(1-\tilde{n})^2}{(1+\tilde{n})^2}\right|\quad.$$
 The refractive index $\tilde{n}$ and the complex dielectric constant $\epsilon=\epsilon_R+i\epsilon_I$, conveying the contribution of conduction electrons, are related\cite{bor} by
  	 $$\tilde{n}^2=\frac{\epsilon}{\epsilon_0}=1-\frac{\left(\omega_p/\omega\right)^2}{1-i/\left(\omega\tau\right)}\quad.$$
At last we get
$$\lambda_L=\frac{2}{\mu_0c\sigma A}\quad,\quad\tau=\mu_0\sigma\lambda_L^2\quad,$$
where $c$ refers to light velocity in vacuum. The same procedure could be applied in normal metals too; however due to $\tau\approx 10^{-14}s$, the available frequency range would be much narrower : $\left[10^{14}Hz,10^{16}Hz\right]$ versus $\left[10^{11}Hz,10^{16}Hz\right]$ in a superconductor.
	\section{conclusion}
	This explanation of the Meissner effect resorts solely to macroscopic arguments. The applied, time-dependent magnetic field excites transient eddy currents according to Newton and Maxwell's equations, which turn to persistent ones, after the magnetic field stops varying and the induced electric field thereby vanishes. Those eddy currents thwart the magnetic field penetration. Were the same experiment to be carried out in a normal metal, eddy currents would have built up the same way. However, once the electric field vanishes, they would have been destroyed quickly by Joule dissipation and the magnetic field would have subsequently penetrated into bulk matter. As a matter of fact, the Meissner effect shows up as a mere outcome of \textit{persistent currents}, the very signature of superconductivity. It is thence unrelated to any microscopic property of the superconducting wave-function\cite{par,bar,coo,pip1,pip,gin,gork}. The common physical significance of the Meissner and skin effects, both stemming from $\epsilon_R(\omega)<0$ for $\omega<\omega_p$, has been unveiled too. A Hall effect has been predicted. Hirsch's prediction\cite{hir,hir2}, regarding an electron charge build up at the outer edge of a superconducting sample, has been confirmed quantitatively.
	\acknowledgments
We thank M. Caffarel, E. Ilisca and G. Waysand for encouragement.


\begin{thebibliography}{00}
\bibitem{not2}
the electric polarization $\alpha E_\theta$ of the electrons bound to the nucleus ($\alpha$ refers to their polarisability coefficient) has been omitted for simplicity but can be taken into account by replacing hereafter $\epsilon_0$ by $\epsilon_0+\alpha$
\end{thebibliography}
\end{document}